\documentclass[letter, longauth]{aa} 

\usepackage{graphicx}
\usepackage{txfonts}

\usepackage[pdfpagelabels=false]{hyperref}      
\hypersetup{colorlinks=true,linkcolor=blue,citecolor=blue,filecolor=blue,urlcolor=blue,}
\usepackage{orcidlink}  

\usepackage{xspace}     

\newcommand*{\hcnone}{\ensuremath{\mathrm{HCN(1-0)}}\xspace} 
\newcommand*{\hcopone}{\ensuremath{\text{HCO}^+\mathrm{(1-0)}}\xspace} 
\newcommand*{\hncone}{\ensuremath{\mathrm{HNC(1-0)}}\xspace} 
\newcommand*{\cstwo}{\ensuremath{\mathrm{CS(2-1)}}\xspace} 
\newcommand*{\coone}{\ensuremath{\mathrm{CO(1-0)}}\xspace} 
\newcommand*{\cotwo}{\ensuremath{\mathrm{CO(2-1)}}\xspace} 
\newcommand*{\htwo}{\ensuremath{\text{H}_2}\xspace} 
\newcommand*{\hone}{\ensuremath{\text{H}\,{\scriptstyle\text{I}}}\xspace}  

\newcommand*{\intCO}{\ensuremath{W_{\text{CO}}}\xspace}  
\newcommand*{\intHCN}{\ensuremath{W_{\text{HCN}}}\xspace}  
\newcommand*{\lco}{\ensuremath{L_{\text{CO}}}\xspace}  
\newcommand*{\lhcn}{\ensuremath{L_{\text{HCN}}}\xspace}  
\newcommand*{\sigmol}{\ensuremath{\Sigma_{\text{mol}}}\xspace}  
\newcommand*{\pde}{\ensuremath{P_{\text{DE}}}\xspace}  
\newcommand*{\alphaco}{\ensuremath{\alpha_{\text{CO}}}\xspace}  
\newcommand*{\alphahcn}{\ensuremath{\alpha_{\text{HCN}}}\xspace}  
\newcommand*{\cir}{\ensuremath{C_\mathrm{IR}}\xspace} 
\newcommand*{\sigsfr}{\ensuremath{\Sigma_{\text{SFR}}}\xspace}  
\newcommand*{\mmol}{\ensuremath{M_{\text{mol}}}\xspace}  
\newcommand*{\mdense}{\ensuremath{M_{\text{dense}}}\xspace}  
\newcommand*{\mstar}{\ensuremath{M_{\star}}\xspace}  
\newcommand*{\sigdense}{\ensuremath{\Sigma_{\text{dense}}}\xspace}  
\newcommand*{\fdense}{\ensuremath{f_{\text{dense}}}\xspace}  
\newcommand*{\sfedense}{\ensuremath{\text{SFE}_{\text{dense}}}\xspace}  
\newcommand*{\taumol}{\ensuremath{\tau^{\text{mol}}_{\rm dep}}\xspace}  
\newcommand*{\taudense}{\ensuremath{\tau_\mathrm{dep}^\mathrm{dense}}\xspace} 
\newcommand*{\tffdense}{\ensuremath{t_\mathrm{ff}^\mathrm{dense}}\xspace} 
\newcommand*{\effdense}{\ensuremath{\epsilon_\mathrm{ff}^\mathrm{dense}}\xspace} 
\newcommand*{\sigstar}{\ensuremath{\Sigma_{\star}}\xspace}  
\newcommand*{\rhostar}{\ensuremath{\rho_{\star}}\xspace}  
\newcommand*{\sigatom}{\ensuremath{\Sigma_{\text{atom}}}\xspace}  
\newcommand*{\siggas}{\ensuremath{\Sigma_{\text{gas}}}\xspace}  

\newcommand*{\ndense}{\ensuremath{n_\mathrm{\htwo}^\mathrm{dense}}\xspace} 

\newcommand*{\rgal}{\ensuremath{r_{\rm gal}}\xspace} 
\newcommand*{\snr}{\ensuremath{\mathrm{S}/\mathrm{N}}\xspace}  

\usepackage{siunitx}
\usepackage{xparse}  
\DeclareSIUnit \parsec {pc}  
\DeclareSIUnit \micron {\micro\metre}  
\DeclareSIUnit \year {yr}  
\DeclareSIUnit \jansky {Jy}  
\DeclareSIUnit \msun {M_{\odot}}  
\DeclareSIUnit \lsun {L_{\odot}}  
\DeclareSIUnit \zsun {Z_{\odot}}  
\DeclareSIUnit \kkms {\kelvin\km\per\second}  
\DeclareSIUnit \kB {\textit{k}_B}  
\DeclareSIUnit \dex {dex}  
\DeclareSIUnit \erg {erg}  
\NewDocumentCommand\angRange{O{} m m}{\SIrange[parse-numbers=false, #1]{\ang[parse-numbers=true]{#2}}{\ang[parse-numbers=true]{#3}}{}}  
\newcommand*{\ra}[2][]{{
                \ang[
                angle-symbol-degree=\textsuperscript{h},
                angle-symbol-minute=\textsuperscript{m},
                angle-symbol-second=\textsuperscript{s},
                #1]{#2}%
}}

\usepackage{pifont}
\newcommand{\cmark}{\textcolor{green}{\ding{51}}}
\newcommand{\xmark}{\textcolor{red}{\ding{55}}}

\usepackage{multirow}  
\usepackage{placeins}  

\begin{document} 

\title{Dense gas scaling relations at kiloparsec scales across nearby galaxies with the ALMA ALMOND and IRAM 30m EMPIRE surveys}
\titlerunning{Dense gas tracer scaling relations}

\author{
Lukas~Neumann\orcidlink{0000-0001-9793-6400}\inst{\ref{aifa},\ref{eso}}\thanks{lukas.neumann.astro@gmail.com};
Mar\'ia~J.~Jim\'enez-Donaire\orcidlink{0000-0002-9165-8080}\inst{\ref{oan},\ref{yebes}};
Frank~Bigiel\orcidlink{0000-0003-0166-9745}\inst{\ref{aifa}};
Adam~K.~Leroy\orcidlink{0000-0002-2545-1700}\inst{\ref{ohio}};
Frank~Bigiel\orcidlink{0000-0003-0166-9745}\inst{\ref{aifa}};
Antonio~Usero\orcidlink{0000-0003-1242-505X}\inst{\ref{oan}};
Jiayi~Sun\orcidlink{0000-0003-0378-4667}\inst{\ref{princeton}};
Eva~Schinnerer\orcidlink{0000-0002-3933-7677}\inst{\ref{mpia}};
Miguel~Querejeta\orcidlink{0000-0002-0472-1011}\inst{\ref{oan}};
Sophia~K.~Stuber\orcidlink{0000-0002-9333-387X}\inst{\ref{mpia}};
Ivana~Be\v{s}li\'c\orcidlink{0000-0003-0583-7363}\inst{\ref{lerma}};
Ashley~Barnes\orcidlink{0000-0003-0410-4504}\inst{\ref{eso}};
Jakob~den~Brok\orcidlink{0000-0002-8760-6157}\inst{\ref{cfa}};
Yixian~Cao\orcidlink{0000-0001-5301-1326}\inst{\ref{mpe}};
Cosima~Eibensteiner\orcidlink{0000-0002-1185-2810}\inst{\ref{nrao}}\thanks{Jansky Fellow of the National Radio Astronomy Observatory};
Hao~He\orcidlink{0000-0001-9020-1858}\inst{\ref{aifa}};
Ralf~S.~Klessen\orcidlink{0000-0002-0560-3172}\inst{\ref{zfa},\ref{iwr},\ref{cfa},\ref{harvard_cashin}};
Fu-Heng~Liang\orcidlink{0000-0003-2496-1247}\inst{\ref{ari}};
Daizhong~Liu\orcidlink{0000-0001-9773-7479}\inst{\ref{purple_mountain}};
Hsi-An~Pan\orcidlink{0000-0002-1370-6964}\inst{\ref{tamkang}};
Thomas~G.~Williams\orcidlink{0000-0002-0012-2142}\inst{\ref{oxford}};
}{}

\authorrunning{Neumann, Jim\'enez-Donaire et al.}

\institute{
Argelander-Institut für Astronomie, Universität Bonn, Auf dem Hügel 71, 53121 Bonn, Germany \label{aifa} \and
European Southern Observatory, Karl-Schwarzschild Stra{\ss}e 2, D-85748 Garching bei M\"{u}nchen, Germany \label{eso} \and
Observatorio Astron\'omico Nacional (IGN), C/ Alfonso XII, 3, E-28014 Madrid, Spain \label{oan} \and
Centro de Desarrollos Tecnol\'ogicos, Observatorio de Yebes (IGN), 19141 Yebes, Guadalajara, Spain \label{yebes} \and
Department of Astronomy, The Ohio State University, 140 West 18th Ave, Columbus, OH 43210, USA \label{ohio} \and
Department of Astrophysical Sciences, Princeton University, 4 Ivy Lane, Princeton, NJ 08544, USA \label{princeton} \and
Max Planck Institute for Astronomy, Königstuhl 17, 69117 Heidelberg, Germany \label{mpia} \and
LERMA, Observatoire de Paris, PSL Research University, CNRS, Sorbonne Universit\'es, 75014 Paris, France \label{lerma} \and
Center for Astrophysics $\mid$ Harvard \& Smithsonian, 60 Garden St., 02138 Cambridge, MA, USA \label{cfa} \and
Max-Planck-Institut f\"ur Extraterrestrische Physik (MPE), Giessenbachstr. 1, D-85748 Garching, Germany \label{mpe} \and
National Radio Astronomy Observatory, 520 Edgemont Road, Charlottesville, VA 22903, USA \label{nrao} \and
Universit\"{a}t Heidelberg, Zentrum f\"{u}r Astronomie, Institut f\"{u}r Theoretische Astrophysik, Albert-Ueberle-Str.\ 2, 69120 Heidelberg, Germany \label{zfa} \and
Universit\"{a}t Heidelberg, Interdisziplin\"{a}res Zentrum f\"{u}r Wissenschaftliches Rechnen, Im Neuenheimer Feld 225, 69120 Heidelberg, Germany \label{iwr} \and
Elizabeth S. and Richard M. Cashin Fellow at the Radcliffe Institute for Advanced Studies at Harvard University, 10 Garden Street, Cambridge, MA 02138, USA \label{harvard_cashin} \and
Astronomisches Rechen-Institut, Zentrum f\"{u}r Astronomie der Universit\"{a}t Heidelberg, M\"{o}nchhofstra\ss e 12-14, D-69120 Heidelberg, Germany \label{ari} \and
Purple Mountain Observatory, Chinese Academy of Sciences, 10 Yuanhua Road, Nanjing 210023, China \label{purple_mountain} \and
Department of Physics, Tamkang University, No.151, Yingzhuan Road, Tamsui District, New Taipei City 251301, Taiwan \label{tamkang} \and
Sub-department of Astrophysics, Department of Physics, University of Oxford, Keble Road, Oxford OX1 3RH, UK \label{oxford}
}

\date{Received 28 November, 2024; accepted 13 December, 2024}

\abstract{
Dense, cold gas is the key ingredient for star formation.
Over the last two decades, HCN(1--0) emission has been the most accessible dense gas tracer for studying external galaxies.
We present new measurements that demonstrate the relationship between dense gas tracers, bulk molecular gas tracers, and star formation in the ALMA ALMOND survey, the largest sample of resolved ($1-2$ kpc resolution) HCN maps of galaxies in the local Universe ($d < 25\,$Mpc). 
We measured HCN/CO, a line ratio sensitive to the physical density distribution, and the star formation rate to HCN ratio (SFR/HCN), a proxy for the dense gas star formation efficiency, as a function of molecular gas surface density, stellar mass surface density, and dynamical equilibrium pressure across 31 galaxies (a factor of $>3$ more compared to the previously largest such study, EMPIRE).
HCN/CO increases (slope of $\approx\num{0.5}$ and scatter of $\approx\SI{0.2}{\dex}$) and SFR/HCN decreases (slope of $\approx\num{-0.6}$ and scatter of $\approx\SI{0.4}{\dex}$) with increasing molecular gas surface density, stellar mass surface density, and pressure. 
Galaxy centres with high stellar mass surface densities show a factor of a few higher HCN/CO and lower SFR/HCN compared to the disc average, but the two environments follow the same average trend.
Our results emphasise that molecular gas properties vary systematically with the galactic environment and demonstrate that the scatter in the Gao--Solomon relation (SFR/HCN) has a physical origin.
}

\keywords{ISM: molecules --
         Galaxies: ISM --
         Galaxies: star formation
        }

\maketitle
%

\section{Introduction}

Stars form from the coldest, densest substructures within molecular clouds. 
Higher-critical-density molecular lines (`dense gas tracers') trace this denser subset of molecular gas, in contrast to the less dense gas probed by low-J CO lines.
The brightest and most commonly used extragalactic dense gas tracers are \hcnone and \hcopone, hereafter HCN and HCO$^+$. A nearly linear correlation has been observed between the star formation rate (SFR) and dense gas tracer luminosity across a wide range of scales \citep[e.g.][]{Gao2004, Wu2010, Garcia-Burillo2012, Usero2015, Chen2017}. This has been interpreted as an indication that dense gas plays a regulating role in the star formation process. This prompts the question of what determines the amount of dense gas. Moreover, these early studies suggested that the rate of star formation per unit dense gas tracer luminosity or dense gas mass (SFE$_{\rm dense} \equiv {\rm SFR}/M_{\rm dense}$) is not universal, but varies from galaxy to galaxy and location to location \citep[][]{Garcia-Burillo2012, Usero2015, Chen2015}.

Recently, the first dense gas tracer mapping surveys that cover whole galaxies have emerged. The Institut de Radioastronomie Millim{\'e}trique (IRAM) 30m large programme EMPIRE\footnote{Eight MIxing Receiver (EMIR) Multiline Probe of the Interstellar medium (ISM) Regulating galaxy Evolution; \url{https://empiresurvey.yourwebsitespace.com}} \citep[][]{Bigiel2016, Jimenez-Donaire2017, Jimenez-Donaire2019} obtained approximately kiloparsec-resolution maps of dense gas tracers (HCN, HCO$^+$, and \hncone) and CO isotopologues for nine nearby galaxies. The ALMA ALMOND\footnote{ACA Large-sample Mapping Of Nearby galaxies in Dense gas.} survey \citep{Neumann2023a} used the Morita Atacama Compact Array (ACA) to map HCN, HCO$^+$, and \cstwo emission from 25 nearby galaxies that were also mapped by the Physics at High Angular resolution in Nearby GalaxieS (PHANGS)--ALMA \cotwo survey \citep{Leroy2021a}. 
Meanwhile, a number of smaller surveys observed dense gas tracers in samples of one to four galaxies \citep[e.g.][]{Tan2018, Gallagher2018a, Gallagher2018b, Querejeta2019, Beslic2021, Heyer2022, Neumann2024, Lin2024}. 

These mapping surveys confirmed significant variations in the SFR/HCN and HCN/CO ratios.
SFR/HCN serves as a proxy for the star formation efficiency in denser gas.
HCN/CO contrasts high and low critical gas density tracers and so probes the physical density distribution. 
Both quantities correlate with the local stellar and gas surface density, dynamical equilibrium pressure, and other environmental factors such that denser gas (higher HCN/CO) and lower SFR/HCN is found in high-surface-density, high-pressure regions \citep[e.g.][]{Jimenez-Donaire2019}.
At face value, this implies an environment-dependent dense gas star formation efficiency and gas density across the discs of star-forming galaxies.

This Letter presents new measurements of HCN/CO and SFR/HCN as a function of local conditions for the 25 galaxies in the ALMA ALMOND survey, which is the largest mapping survey of dense gas tracers in galaxy discs. 
We synthesised ALMOND data with the IRAM 30m EMPIRE survey, the first large dense gas tracer mapping survey, to present a homogeneously measured set of kiloparsec-resolution scaling relations that connect star formation, dense gas, and total molecular gas to these local environmental quantities for a total of 31 local spiral galaxies. 

\section{Data and methods}
\label{sec:data}
We used the HCN data presented in \citet[][EMPIRE]{Jimenez-Donaire2019} and \citet[][ALMOND]{Neumann2023a}.
The two datasets are well matched in terms of sensitivity and resolution (see Appendix~\ref{sec:app:empire_vs_almond}; median 1.47~kpc, 16{-}84\% range 1.09{-}1.76 kpc). 
We convolved all supporting data to the angular resolution of the HCN observations using the \href{https://zenodo.org/records/13787728}{PyStructure} package\footnote{\url{https://github.com/jdenbrok/PyStructure}}. 
We sampled all maps with a half-beam-spaced hexagonal grid and computed the integrated intensities of the HCN and CO lines by integrating over a velocity range determined by the velocity extent of the CO emission. 
The velocity integration mask was built from 4-sigma CO peaks and expanded into adjacent 2-sigma channels.
We treated the ratios HCN/CO and SFR/HCN as the quantities of interest and measured how they vary as a function of the molecular gas mass surface density ($\Sigma_{\rm mol}$), stellar mass surface density ($\Sigma_\star$), and dynamical equilibrium pressure ($P_{\rm DE}$).

\paragraph{Galaxy sample.} 

\begin{figure}
\centering
\includegraphics[width=\columnwidth]{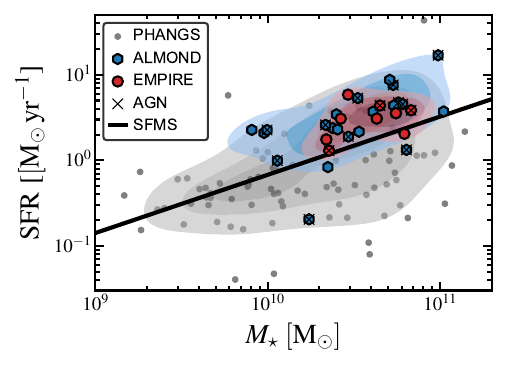}
\caption{ALMOND and EMPIRE on the star-forming main sequence (SFMS) of galaxies.
Grey shows all galaxies from the PHANGS--ALMA survey \citep{Leroy2021b}.
Red and blue markers present galaxies from the EMPIRE \citep{Jimenez-Donaire2019} and ALMOND surveys \citep{Neumann2023a}, respectively, with the same SFR calibration adopted across the merged sample.
Contours indicate 25, 50, and 75 percentile areas of the respective samples.
The solid black line marks the star-forming main sequence from z0MGS \citep{Leroy2019}.
The black squares and crosses indicate the presence of a bar or AGN in the respective galaxy, taken from Table~\ref{tab:sample}.
}
\label{fig:sfms}
\end{figure}

Table~\ref{tab:sample} lists our targets, their integrated properties, survey coverage, and the resolution of the HCN observations. 
ALMOND and EMPIRE target nearby ($d<25\,\mathrm{Mpc}$), $i<75^\circ$, star-forming ($\mathrm{SFR}\sim0.2-17\,M_\odot\,\mathrm{yr}^{-1}$) galaxies, which span stellar masses from \SI{8e9}{\msun} to \SI{1e11}{\msun} and SFRs from \SI{0.2}{\msun\per\year} to \SI{17}{\msun\per\year}.
We stress that ALMOND significantly increases the dynamic range of SFR and \mstar (by approximately a factor of 2) compared to the previous largest sample (i.e. EMPIRE; see Fig.~\ref{fig:sfms}).
Combining the surveys yields 31 unique galaxies.
NGC\,628, 2903, and 4321 were observed by both surveys, and their measurements are consistent (see Appendix~\ref{sec:app:empire_vs_almond}). 
To avoid duplicates, we employed the ALMOND data for these galaxies.

\paragraph{Star formation rate.}
We estimate the kiloparsec-scale SFR and SFR surface density ($\Sigma_{\rm SFR}$) following the methodology of the original ALMOND paper \citep{Neumann2023a}, which used a combination of infrared (IR; \SI{22}{\micron}) maps from Wide-field Infrared Survey Explore \citep[WISE;][]{Wright2010} and far-ultraviolet (FUV; \SI{154}{nm}) maps from Galaxy Evolution Explorer \citep[GALEX;][]{Martin2005}. These maps were taken from the z0MGS atlas \citep{Leroy2019} and converted to SFR following their best FUV+22$\mu$m prescription. Although EMPIRE employed \textit{Spitzer} and \textit{Herschel} IR measurements to estimate the SFR, here we adopted the same methodology across EMPIRE and ALMOND, using the FUV+22$\mu$m-based SFR maps for the full sample.

\paragraph{Stellar mass.}
We estimated the stellar mass surface density (\sigstar) from \textit{Spitzer} \SI{3.6}{\micron} observations \citep{Sheth2010,Querejeta2021b}. We used the dust-corrected maps from \citet{Querejeta2015} and adopted a mass-to-light ratio of $\Upsilon_\star = 0.6$~M$_\odot$~L$_\odot^{-1}$.

\paragraph{Dynamical equilibrium pressure.}
The dynamical equilibrium pressure (\pde) expresses the total interstellar pressure needed to support a disc in vertical dynamical equilibrium \citep[e.g. see][]{Ostriker2022, Schinnerer2024}. We estimated \pde by calculating the weight of the interstellar medium in the galaxy potential via\begin{align}
    \pde = \dfrac{\pi G}{2}\siggas^2 + \siggas \, \sqrt{2G\rhostar} \, \sigma_{\rm gas, z} \;,
\end{align}
where $\siggas=\sigmol + \sigatom$ is the total gas surface density, $\rhostar$ is the stellar mass volume density, and $\sigma_{\rm gas, z}=\SI{15}{\km\per\s}$ \citep[e.g.][]{Sun2018} is the gas velocity dispersion perpendicular to the galactic disc.
Computing \rhostar required estimates of the stellar scale heights, which were estimated from measured stellar disc scale lengths by assuming a typical disc flattening ratio \citep[see][for more details]{Sun2020b,Sun2022}.
Estimating \pde additionally required measurements of the atomic gas content, which were taken from \hone 21\,cm line observations (all with angular resolutions similar to or higher  than that of HCN). These observations were available for 26 galaxies of our sample (Table~\ref{tab:sample}), hence limiting the analysis of \pde relations to those 26 galaxies.

\begin{figure*}
\centering
\includegraphics[width=\textwidth]{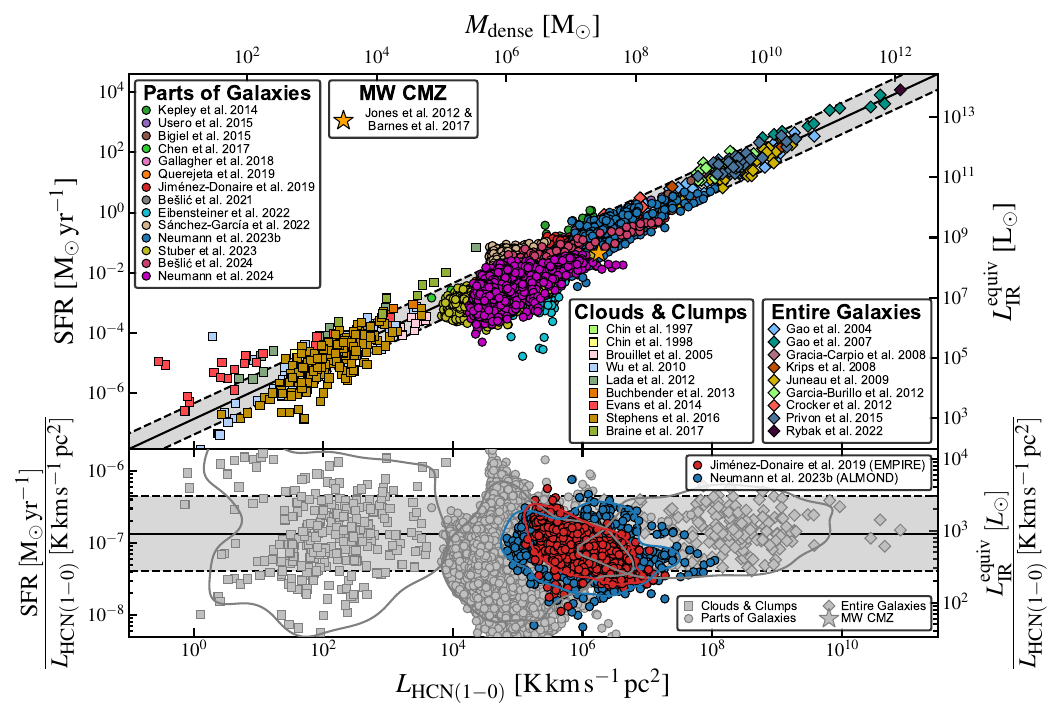}
\caption{Gao--Solomon relation. SFR (top) and $\mathrm{SFR}/\lhcn$ (a proxy of \sfedense; bottom) as a function of HCN luminosity across a literature compilation and the ALMOND (blue circles) and EMPIRE (red circles) surveys.
Note that we re-calculated the SFR across EMPIRE galaxies using a combination of IR and FUV data (see Sect.~\ref{sec:data}).
Our literature compilation contains HCN observations that include Galactic clumps and clouds (squares), resolved nearby galaxies (circles), and unresolved entire galaxies (diamonds).
For more details on the compilation, see Appendix~\ref{sec:app:literature}.
The plotted data points show all (3-sigma) detected sightlines.
The solid black line shows the median SFR/HCN computed from these data points across all datasets (without duplicates across targets), and the dashed lines mark the 1-sigma scatter (Table~\ref{tab:gao_solomon}).
The bottom panel shows the ratio SFR/HCN as a function of \lhcn, grouping the data into the same subsamples, for which the 10-percentile density contours of the respective subsamples are shown.
We plot ALMOND and EMPIRE data separately, and the blue and red contours present the 10-percentile levels of these surveys.
}
\label{fig:gao_solomon_relation}
\end{figure*}

\begin{table*}
    \begin{center}
    \caption{Gao--Solomon relation.}
    \label{tab:gao_solomon}
    \begin{tabular}{ccccc}
    \hline \hline
    \multirow{3}{*}{Regime} & $\log_{10}\,$SFR/HCN & $\log_{10}\,$IR/HCN & $\log_{10}\,\taudense$  & $\sigma$ \\
     & [$\SI{}{\msun\per\year}/(\SI{}{\kkms\square\parsec})$] & [$\SI{}{\lsun}/(\SI{}{\kkms\square\parsec})$] & [$\SI{}{\year}$] & $[\SI{}{\dex}]$ \\
     & (16$^\mathrm{th}$, 50$^\mathrm{th}$, 84$^\mathrm{th}$) perc. & (16$^\mathrm{th}$, 50$^\mathrm{th}$, 84$^\mathrm{th}$) perc. & (16$^\mathrm{th}$, 50$^\mathrm{th}$, 84$^\mathrm{th}$) perc. &  \\
     \hline
    Clouds \& Clumps  & $(-7.43,-6.89,-6.27)$ & $(2.40,2.94,3.56)$ & $(7.45,8.07,8.61)$ & $0.70$ \\
    Parts of Galaxies & $(-7.23,-6.98,-6.65)$ & $(2.60,2.85,3.18)$ & $(7.82,8.16,8.41)$ & $0.46$ \\
    Entire Galaxies   & $(-7.16,-6.85,-6.56)$ & $(2.67,2.98,3.27)$ & $(7.74,8.03,8.33)$ & $0.27$ \\
    Combined          & $(-7.17,-6.87,-6.48)$ & $(2.66,2.96,3.35)$ & $(7.66,8.05,8.34)$ & $0.52$ \\
    ALMOND \& EMPIRE  & $(-7.14,-6.84,-6.44)$ & $(2.69,2.99,3.39)$ & $(7.62,8.02,8.31)$ & $0.35$ \\
    \hline\hline
    \end{tabular}
    \end{center}
    {\bf Notes} -- Median dense gas ratios across the combined literature sample presented in Fig.~\ref{fig:gao_solomon_relation}, including ALMOND and EMPIRE, and for respective subsamples, including clouds and clumps and resolved and integrated galaxy surveys.
    The values across the `Parts of Galaxies' studies are computed from unique targets \citep[i.e.][to avoid target duplication]{Bigiel2015, Kepley2014, Jimenez-Donaire2019, Sanchez-Garcia2022, Neumann2023a, Beslic2024}.
    The `Combined' measurements are computed from the medians of each respective study.
    `ALMOND \& EMPIRE' results were obtained from the medians of each respective galaxy (i.e. from 31 galaxies) and consider non-detections (in contrast to the other columns, which only consider 3-sigma detected data).
    All values are on a logarithmic scale.
    Columns 2 and 3 list the 16th percentiles, medians, and 84th percentiles of $\log_{10}\,$SFR/HCN and $\log_{10}\,$IR/HCN. 
    Column 4 displays the dense gas depletion time (\taudense) in units of years, and Col. 5 the 1-sigma scatter ($\sigma$) of the detected HCN data around the median value in units of dex.
    See Appendix~\ref{sec:app:literature} for details of the compilation.
\end{table*}

\paragraph{Conversion factors, molecular gas surface density, and dense gas fraction.} 
We focused on the ratios HCN/CO and SFR/HCN. 
For reference, we converted them to fiducial physical quantities using fixed conversion factors, $\alphaco\equiv\mmol/\lco\equiv\sigmol/\intCO$ and $\alphahcn\equiv\mdense/\lhcn\equiv\sigdense/\intHCN$, adopting $\alphaco^{\rm fix}=\SI{4.35}{\msun\per\square\parsec}\,(\SI{}{\kkms})^{-1}$ \citep{Bolatto2013} and $\alphahcn=\SI{15}{\msun\per\square\parsec}\,(\SI{}{\kkms})^{-1}$ \citep{Schinnerer2024}. 
Aside from aiming to remain `close to the observations', we did this because the environmental dependence of the HCN-to-dense gas conversion factor, $\alpha_{\rm HCN}$, remains unclear, with no obvious best prescription and likely significant covariance with $\alpha_{\rm CO}$ \citep[see][]{Usero2015}.

Nevertheless, to leverage recent progress in understanding \alphaco variations, we employed a variable \alphaco (hereafter $\alphaco^{\rm var}$) when considering molecular gas surface density, \sigmol, or dynamical equilibrium pressure (\pde; see the paragraphs below) as independent variables (i.e. on the $x$-axis). We calculated
\begin{align}
\left(\dfrac{\sigmol}{\SI{}{\msun\per\square\parsec}}\right) = \alphaco^{\rm var}\,\left(\dfrac{\intCO}{\SI{}{\kkms}}\right) \cos(i) \;,
\end{align}
adopting the $\alphaco^{\rm var}$ prescription from \citet{Schinnerer2024}. This $\alphaco^{\rm var}$ depends on metallicity and $\Sigma_\star$ (see Appendix~\ref{sec:app:conversion_factors} for more details on the variable conversion factor and line ratio prescriptions, including references to the works synthesised by \citealt{Schinnerer2024}).

When quoting HCN/CO, we cast our results in terms of \coone and employed the $\cotwo/\coone$ line ratio calibration as a function of \sigsfr to convert PHANGS--ALMA \cotwo to \coone intensities (see Appendix~\ref{sec:app:conversion_factors}). 
EMPIRE already has \coone maps.

For both quantities, we provide reference conversions to physical units. We calculated the `dense gas fraction' as the ratio between dense and bulk molecular gas using fixed conversion factors, $\fdense \equiv \mdense/\mmol \propto {\rm HCN/CO}$, as 
\begin{align}
\fdense \approx \num{3.5}\,\left(\dfrac{\intHCN}{\SI{}{\kkms}}\right)\left(\dfrac{\intCO}{\SI{}{\kkms}}\right)^{-1} \;.
\end{align}

\noindent We converted SFR/HCN to an approximate star formation efficiency of dense molecular gas, $\sfedense\equiv \rm{SFR}/\mdense$, via\begin{align}
\left(\dfrac{\sfedense}{\SI{}{\per\year}}\right) \approx \num{6.7e-2}\,\left(\dfrac{\sigsfr}{\SI{}{\msun\per\year\per\square\parsec}}\right)\left(\dfrac{\intHCN}{\SI{}{\kkms}}\right)^{-1} \;.
\end{align}

\section{Results and discussion}
\label{sec:results}

\subsection{Gao--Solomon relation}

In Fig.~\ref{fig:gao_solomon_relation} we present the `Gao--Solomon' relation \citep{Gao2004}, the scaling relationship between SFR and $L_{\rm HCN}$. We placed ALMOND and EMPIRE in the context of a literature compilation that comprises 31 HCN surveys spanning from the Milky Way to the high-redshift universe. This includes observations of individual cores and molecular clouds within the Milky Way and the Local Group, spatially resolved maps of galaxies, and integrated galaxy data. On the $x$- and $y$-axes, we indicate both the observed luminosities (HCN and IR) and the inferred physical quantities (\mdense and SFR), assuming linear conversions with fixed conversion factors \alphahcn and \cir\footnote{All data here have observed HCN, but for the $y$-axis we adopted the best-estimate SFR and converted it to the equivalent $L_{\rm IR}$ using a constant IR-to-SFR conversion factor, $\cir=\SI{1.48e-10}{\msun\per\year\per\lsun}$ \citep{Murphy2011}.}. ALMOND and EMPIRE form the largest resolved galaxy dataset, filling in the large gap in spatial scale, SFR, and $L_{\rm HCN}$ between integrated galaxy and individual cloud studies.

In the bottom panel of Fig.~\ref{fig:gao_solomon_relation}, the $y$-axis displays the ratio between SFR and \lhcn. 
Across the full literature sample, we find a median SFR/HCN of $\SI{1.3e-07}{\msun\per\year}\,(\SI{}{\kkms\square\parsec})^{-1}$ with a 1-sigma scatter of $\SI{0.52}{\dex}$, which is consistent with previous literature compilations \citep[e.g.][]{Jimenez-Donaire2019, Beslic2024}.
We also computed the respective median \sfedense values and scatter for the individual subsamples: clumps and clouds (squares), resolved galaxy observations (circles), and entire galaxies (diamonds). 
The values, along with the values specifically for ALMOND and EMPIRE, are listed in Table~\ref{tab:gao_solomon}.
Overall, the literature compilation demonstrates that the HCN luminosity is a reasonable predictor of the SFR from cloud to galaxy scales across ten orders of magnitude. 
However, at any given HCN luminosity, there is a significant scatter, $\sigma\sim\SI{0.5}{\dex}$.
Moreover, the scatter increases from large (entire galaxy; $\sigma=\SI{0.27}{\dex}$) to small scales (clouds; $\sigma=\SI{0.70}{\dex}$), suggesting that there are significant variations in \sfedense within galaxies (discussed in Sect.~\ref{sec:environment_relations}) that average out at integrated galaxy scales.

SFR/HCN can be interpreted, with significant uncertainty due to the uncertain conversion factor, as the rate per unit mass at which dense molecular gas converts into stars. Across the detected sightlines of the full literature sample, we find a median $\sfedense \approx \SI{8.9e-09}{\per\year}$, or equivalently a median dense gas depletion time of $\taudense \approx \SI{112}{\mega\year}$, indicating that the rate of present-day star formation would consume the available dense gas in this time period. 
For reference, this is $\approx 10$ times lower than estimates for \taumol, the overall molecular gas depletion time in similar samples \citep[][]{Sun2023}. 
The star formation efficiency per freefall time, $\effdense=\sfedense\cdot\tffdense$, is of theoretical interest \citep[e.g.][]{Krumholz2005, Federrath2012} because it captures the efficiency of star formation relative to the timescale expected for gravitational collapse, and so normalises for density.
The freefall time of the dense molecular gas was computed as $\tffdense=\SI{0.8}{\mega\year}$, assuming that HCN traces gas above a density of $\ndense\approx \SI{3e3}{\per\cubic\centi\metre}$ \citep{Jones2023, Bemis2024}.
Across the full literature sample, we obtain a median $\effdense\approx\SI{0.7}{\percent}$, which suggests that only \SI{0.7}{\percent} of the dense molecular gas is converted into stars per gravitational collapse timescale.
This demonstrates that even in the dense gas, star formation appears to be an extremely inefficient process.

\begin{figure*}
\centering
\includegraphics[width=\textwidth]{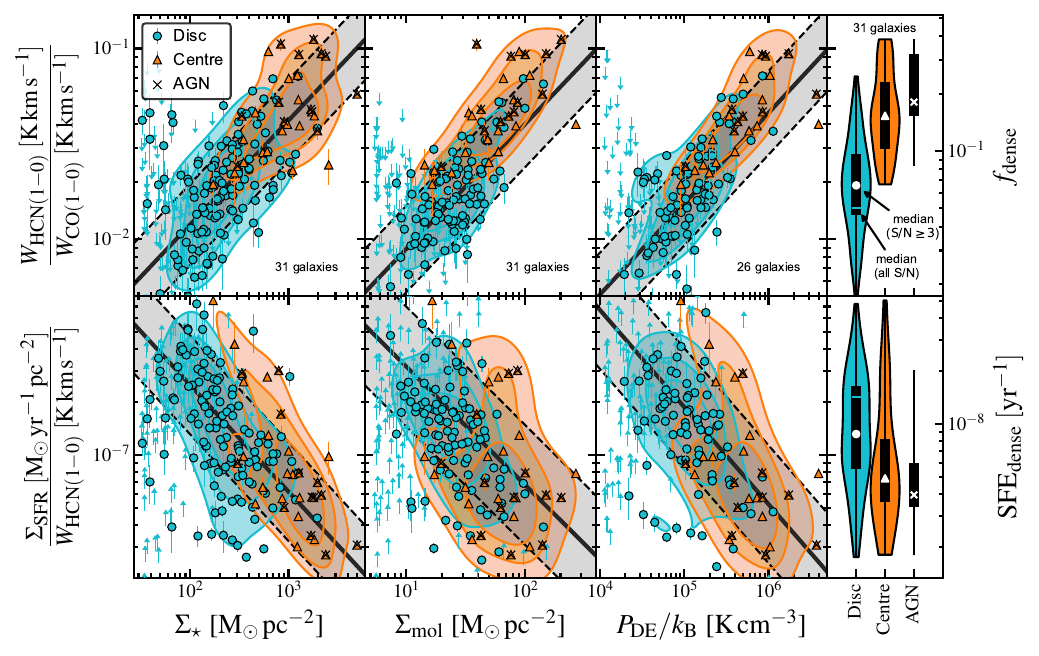}
\caption{Dense gas relations with kiloparsec-scale environmental conditions. 
HCN/CO \textit{(top)}, a proxy of \fdense, and SFR/HCN \textit{(bottom)}, a proxy of \sfedense, are shown as a function of stellar mass surface density (\sigstar), molecular gas surface density (\sigmol), and dynamical equilibrium pressure (\pde) across 31 galaxies from ALMOND and EMPIRE.
The markers denote significant stacked measurements  ($\snr\geq 3$) across disc (circle) and centre (triangle) spaxels.
The downward and upward pointing arrows denote upper (HCN/CO) and lower limits (SFR/HCN).
Filled contours show 25, 50, and 75 percentile kernel density estimates.
Across centres, we indicate the presence of an AGN (cross).
All relations have been fitted with \texttt{LinMix}, taking measurement uncertainties and upper and lower limits into account (parameters in Table~\ref{tab:environment}).
The solid black line shows the best-fit line, and the grey-shaded area indicates the 1-sigma scatter of $\snr\geq 3$ data.
The right panels show violin plots of the HCN/CO and SFR/HCN distribution across the respective samples (disc, centre, centre with an AGN), where the black bar and white markers indicate the 25th to 75th percentile range and the median, respectively, across the $\snr\geq 3$ data.
The vertical cyan lines in the disc violins mark the median computed from all \snr data.
}
\label{fig:hcn_scaling_relations}
\end{figure*}

\subsection{Dense gas relations with environmental conditions}
\label{sec:environment_relations}

Figure \ref{fig:gao_solomon_relation} shows significant scatter in SFR/HCN. Previous works have found that both SFR/HCN and HCN/CO depend systematically on environmental factors, including the stellar mass surface density (\sigstar), the molecular gas mass surface density (\sigmol), and the interstellar pressure inferred from dynamical equilibrium (\pde; \citealt{Usero2015, Gallagher2018a, Jimenez-Donaire2019}). The combined ALMOND and EMPIRE samples are ideal for measuring these environmental variations. 
Individual regions follow the overall Gao--Solomon relation and show a comparable scatter to the full literature sample. The kiloparsec-scale resolution is, on the one hand, high enough to resolve galaxies into discrete regions, including centres, bars, and spiral arms, but is, on the other hand, coarse enough to average over many individual regions to access the time-averaged mean HCN/CO and SFR/HCN.

In Fig.~\ref{fig:hcn_scaling_relations} we use ALMOND and EMPIRE to make the most rigorous measurement to date of the scaling relations relating HCN/CO and SFR/HCN to these environmental factors. For each galaxy we spectrally stacked the \hcnone and \coone lines in bins of \sigstar, \sigmol, and \pde using \texttt{PyStacker}\footnote{\url{https://github.com/PhangsTeam/PyStacker}}. We used the CO data, which have a much higher signal-to-noise ratio than the HCN, to determine the local mean reference velocity for the stacks \citep[see][and references therein for details on the spectral stacking methodology; Appendix~\ref{sec:app:stacking} presents the spectral stacks of HCN and CO]{Neumann2023b}. 
For bins in which the stacks do not yield 3-sigma HCN detections, we estimated upper limits for HCN/CO and lower limits for SFR/HCN. 
We fitted the combined set of stacks (including upper limits) for all galaxies using a linear function of the form
\begin{align}
        \log_{10} Y = b + m \cdot (\log_{10} X - x_0)\;,
    \label{equ:fit_function}
\end{align}
where $X=\{\sigstar, \sigmol, \pde\}$ and $Y=\{\mathrm{HCN/CO}, \mathrm{SFR/HCN}\}$ are the $x$- and $y$-axis variables, respectively. 
The slopes and intercepts are denoted as $m$ and $b$, and $x_0=\{2.4, 1.4, 5.0\}$ is a value close to the median $X$ value. 
Centring the fit at $x_0$ minimises the covariance between $m$ and $b$. The fitting was performed with the linear regression tool \texttt{LinMix}\footnote{\url{https://github.com/jmeyers314/linmix}}, which takes measurement uncertainties and 3-sigma upper (lower) limits on HCN/CO (SFR/HCN) into account \citep[see e.g.][for more details on the fitting routine]{Neumann2023a}.
The fit parameters are presented in Table~\ref{tab:environment}. We note that the range of $\Sigma_\star$, $\Sigma_{\rm mol}$, and $P_{\rm DE}$ covered by these results corresponds to the inner, molecular-gas-dominated parts of galaxies, where most stars form.

We find strong correlations between the stacked HCN/CO and all three quantities, and anti-correlations between SFR/HCN and the same quantities (Fig.~\ref{fig:hcn_scaling_relations} and Table~\ref{tab:environment}). 
HCN/CO increases and SFR/HCN decreases with increasing \sigstar, \sigmol, and \pde. 
The slopes are significant, with both HCN/CO and SFR/HCN changing by $\sim\SI{1}{\dex}$ across our sample. 
ALMOND and EMPIRE have consistent results despite using different telescopes and using different CO lines (see Appendix~\ref{sec:app:empire_vs_almond}).

\begin{table}
\begin{center}
\caption{Dense gas tracer and environment in ALMOND and EMPIRE.}
\label{tab:environment}
\resizebox{\columnwidth}{!}{
\begin{tabular}{ccccccc}
\hline \hline
$\log_{10}(Y)$ & $\log_{10}(X)$ & $x_0$ & $m$ (unc.) & $b$ (unc.) & $\sigma$ & $r_\mathrm{Pearson}$ \\
\hline          
\multirow{3}{*}{HCN/CO} & \sigstar & $2.4$  & $0.55$ $(0.03)$ & $-1.71$ $(0.01)$ & $0.21$ & $0.77$ \\
                        & \sigmol  & $1.4$  & $0.65$ $(0.05)$ & $-1.79$ $(0.02)$ & $0.23$ & $0.70$ \\
                        & $\pde/\SI{}{\kB}$ & $5.0$ & $0.48$ $(0.04)$ & $-1.81$ $(0.02)$ & $0.22$ & $0.76$ \\
\hline
\multirow{3}{*}{SFR/HCN}        & \sigstar & $2.4$  & $-0.61$ $(0.04)$ & $-6.84$ $(0.02)$ & $0.28$ & $-0.70$ \\
                            & \sigmol  & $1.4$  & $-0.67$ $(0.08)$ & $-6.77$ $(0.02)$ & $0.34$ & $-0.56$ \\
                            & $\pde/\SI{}{\kB}$ & $5.0$ & $-0.55$ $(0.05)$ & $-6.73$ $(0.03)$ & $0.34$ & $-0.66$ \\
 \hline\hline
\end{tabular}
}
\end{center}
{\bf Notes} --
Fit parameters obtained via linear regression with \texttt{LinMix}  via Eq.~\ref{equ:fit_function} to the data shown in Fig.~\ref{fig:hcn_scaling_relations}. 
The parameters $x_0$, $m$, $b,$ and $\sigma$ are the $x$-axis offset, slope, intercept, and scatter of the relation.
$r_\mathrm{Pearson}$ denotes the Pearson correlation coefficient; all $p$-values are much less than $0.01$.
\sigstar and \sigmol are given in units of \SI{}{\msun\per\square\parsec}; and $\pde/\SI{}{\kB}$ in \SI{}{\kelvin\per\cubic\centi\metre}.\\
\end{table}

The enhanced HCN/CO in high-surface-density, high-pressure environments indicates that a deeper gravitational potential and more abundant overall molecular gas lead to the formation of denser molecular clouds. 
This picture agrees well with the one that has emerged from high-physical-resolution CO imaging, which shows that the mean cloud-scale gas surface density and velocity dispersion correlate with these same environmental factors \citep[][]{Sun2022}. 
In fact, one of the main results from ALMOND has been a good direct correlation between the cloud-scale gas properties and the density-sensitive HCN/CO line ratio \citep{Neumann2023a}. 
The fact that spectroscopic (presented here) and CO imaging results show similar trends as a function of galactic environment provides strong evidence that the physical properties of molecular clouds vary as a function of the galactic environment. 
The HCN/CO variations that we observe are continuous across the whole range of our sample, with a $\sim\SI{0.2}{\dex}$ scatter about the correlation.

SFR/HCN anti-correlates with \sigstar, \sigmol, and \pde. 
This anti-correlation is also significant, though the correlation coefficient is weaker, and the data show more residual scatter in SFR/HCN compared to the trends in HCN/CO. 
At face value, this indicates that the denser molecular gas that effectively emits HCN is less efficiently converted to stars in high-surface-density, high-pressure parts of galaxies. 
A popular explanation for this trend has been that HCN-emitting material in these denser environments does not necessarily uniquely correspond to the overdense, self-gravitating parts of clouds that collapse to form stars \citep[e.g.][]{Krumholz2007, Shetty2014, Gallagher2018b, Neumann2023a, Bemis2023, Bemis2024}

\subsection{Dense gas ratios in galaxy centres}
\label{sec:centres}

The centres of galaxies often exhibit high \sigmol, \sigstar, and \pde; hence, one expects high HCN/CO and low SFR/HCN in galaxy centres compared to the discs. 
In Fig.~\ref{fig:hcn_scaling_relations} we separately indicate galaxy centres in contrast to the rest of the galaxy.
For this exercise, we considered the central kiloparsec-scale, beam-size aperture as the centre and refer to the remaining galaxy parts as the disc.

We find that centres typically have high HCN/CO (median of $0.045$; see Table~\ref{tab:centres}) compared to the discs (median of $0.013$) that are not consistent within the 1-sigma scatter and low SFR/HCN (median of $\SI{7.8e-8}{\msun\per\year\per\square\parsec}\,(\SI{}{\kkms})^{-1}$ compared to the disc median of $\SI{1.3e-7}{\msun\per\year\per\square\parsec}\,(\SI{}{\kkms})^{-1}$ (but overlapping 1-sigma intervals).
In the following, we base our discussion of the centre-disc comparison on the relations with \sigstar, which have uncorrelated axes, in contrast to the relations with \sigmol and \pde, which depend on the CO line intensity.
To first order, centres appear to follow the same average HCN/CO and SFR/HCN against \sigstar trend, showing a continuous extension of the disc trends.\ The higher HCN/CO (lower SFR/HCN) in galaxy centres then simply results from the high \sigstar (\sigmol, \pde) environment of centres.
However, there are some deviations from this simple picture in the SFR/HCN against \sigstar relation.
On the one hand, the disc measurements in intermediate \sigstar environments ($\sigstar\approx\SI{2e2}{\msun\per\square\parsec}-\SI{1e3}{\msun\per\square\parsec}$) tend to have low SFR/HCN compared to the average trend, while centres show high SFR/HCN across the same \sigstar range.
These deviations could be explained via variations with dynamical environments (e.g. \citealt{Neumann2024} found a low SFR/HCN in the galactic bar) but remain speculative due to the coarse, kiloparsec-scale resolution of the ALMOND and EMPIRE observations and hence require higher-resolution observations that resolve these morphological regions.

If taken at face value, the low \sfedense in galaxy centres could imply that these environments are typically less efficient at forming stars per unit of dense gas mass, which could be explained by the higher gas turbulence in these environments acting against gravitational collapse \citep[e.g.][]{Usero2015, Neumann2023a}.
However, a similarly likely explanation is that HCN might not be a robust tracer of dense gas in galaxy centres -- for example due to increased optical depth, IR pumping (e.g. \citealt{Matsushita2015}), or electron excitation, (e.g. \citealp{Goldsmith2018}) -- and potentially trace more of the bulk gas in these high-density regions (see the explanation above). 
Therefore, we might expect \alphahcn to vary between disc and centre regions.
For instance, if one assumes that \alphahcn variations are driven by optical depth effects and vary similarly to \alphaco \citep{Teng2023, Bemis2024}, \alphahcn would be lower in galaxy centres and thus yield higher \sfedense that are more comparable to disc values.

One might expect that active galactic nuclei (AGNs) boost HCN emission \citep[e.g.][]{Goldsmith2018, Matsushita2015}, deplete gas \citep[e.g.][]{Ellison2021}, or quench star formation \citep[e.g.][]{Nelson2019}.
In Fig.~\ref{fig:hcn_scaling_relations} we additionally indicate the presence of an AGN (cross; 14 galaxies) for the galaxy centres and show their median and distribution in the right panels.
We find that active centres have \SI{50}{\percent} higher HCN/CO and lower SFRs, though distributions are similar to those found in non-active galaxies and the differences are not significant at the 1-sigma level.
The variations in dense gas and star formation in AGN-affected regions are likely not well resolved at the scales probed in this study  ($\sim 1-2\,$kpc) and require sub-kiloparsec-resolution observations.

\section{Conclusions}
\label{sec:conclusions}

We present the resolved $1{-}2$~kpc resolution dense gas tracer scaling relations for ALMA ALMOND, a survey of HCN emission from 25 star-forming disc galaxies. 
Combining ALMOND with the IRAM 30m EMPIRE survey, we measured how HCN/CO and SFR/HCN, observational quantities sensitive to the gas density and star formation efficiency of dense gas, depend on the local stellar and molecular gas mass surface density ($\Sigma_\star$ and $\Sigma_{\rm SFR}$) and the estimated dynamical equilibrium pressure ($P_{\rm DE}$). 
Our total sample of 31 resolved galaxies represents a factor of $>3$ increase in the number of galaxies compared to the previous state-of-the-art dense gas mapping surveys.
HCN/CO correlates with all three environmental measures, showing similar trends to those found for cloud-scale ($\sim 100$~pc) CO imaging. 
Our results support the view that the physical state of molecular gas depends on the galactic environment. SFR/HCN anti-correlates with surface density and $P_{\rm DE}$, though they show moderately more scatter than the HCN/CO correlations. This reinforces the notion that the scatter in the Gao--Solomon relation is physical in origin and that the relation between any specific dense gas tracer and star formation activity is environment-dependent. While the physical explanations for each of these trends remain subjects of active research, their presence in the data is clear, and ALMA ALMOND + IRAM 30m EMPIRE provides the best measurement to date in the molecular-gas-dominated, star-forming parts of massive disc galaxies.

\begin{acknowledgements}
We would like to thank the referee, Mark Heyer, for his constructive and concise feedback that helped improve the paper. 
This work was carried out as part of the PHANGS collaboration.
LN acknowledges funding from the Deutsche Forschungsgemeinschaft (DFG, German Research Foundation) - 516405419.
AKL gratefully acknowledges support by grants 1653300 and 2205628 from the National Science Foundation, by award JWST-GO-02107.009-A, and by a Humboldt Research Award from the Alexander von Humboldt Foundation.
AU acknowledges support from the Spanish grants PGC2018-094671-B-I00, funded by MCIN/AEI/10.13039/501100011033 and by ``ERDF A way of making Europe'', and PID2019-108765GB-I00, funded by MCIN/AEI/10.13039/501100011033.
RSK acknowledges financial support via the ERC Synergy Grant ``ECOGAL'' (project ID 855130), via the Heidelberg Cluster of Excellence (EXC 2181 - 390900948) ``STRUCTURES'', and via the BMWK project ``MAINN'' (funding ID 50OO2206). RSK also thanks the 2024/25 Class of Radcliffe Fellows for highly interesting and stimulating discussions.
FHL gratefully acknowledges funding from the European Research Council’s starting grant ERC StG-101077573 (“ISM-METALS”).
HAP acknowledges support from the National Science and Technology Council of Taiwan under grant 110-2112-M-032-020-MY3.

This paper makes use of the following ALMA data, which have been processed as part of the ALMOND and PHANGS--ALMA surveys: \\
\noindent ADS/JAO.ALMA\#2012.1.00650.S, \linebreak 
ADS/JAO.ALMA\#2013.1.01161.S, \linebreak 
ADS/JAO.ALMA\#2015.1.00925.S, \linebreak 
ADS/JAO.ALMA\#2015.1.00956.S, \linebreak 
ADS/JAO.ALMA\#2017.1.00230.S, \linebreak 
ADS/JAO.ALMA\#2017.1.00392.S, \linebreak 
ADS/JAO.ALMA\#2017.1.00766.S, \linebreak 
ADS/JAO.ALMA\#2017.1.00815.S, \linebreak 
ADS/JAO.ALMA\#2017.1.00886.L, \linebreak 
ADS/JAO.ALMA\#2018.1.01171.S, \linebreak 
ADS/JAO.ALMA\#2018.1.01651.S, \linebreak 
ADS/JAO.ALMA\#2018.A.00062.S. \linebreak 
ADS/JAO.ALMA\#2019.2.00134.S, \linebreak 
ADS/JAO.ALMA\#2021.1.00740.S, \linebreak 
ALMA is a partnership of ESO (representing its member states), NSF (USA), and NINS (Japan), together with NRC (Canada), NSC and ASIAA (Taiwan), and KASI (Republic of Korea), in cooperation with the Republic of Chile. The Joint ALMA Observatory is operated by ESO, AUI/NRAO, and NAOJ. The National Radio Astronomy Observatory (NRAO) is a facility of the National Science Foundation operated under cooperative agreement by Associated Universities, Inc.

This work makes use of data products from the Wide-field Infrared Survey Explorer (WISE), which is a joint project of the University of California, Los Angeles, and the Jet Propulsion Laboratory/California Institute of Technology, funded by NASA.

This work is based in part on observations made with the Galaxy Evolution Explorer (GALEX). GALEX is a NASA Small Explorer, whose mission was developed in cooperation with the Centre National d'Etudes Spatiales (CNES) of France and the Korean Ministry of Science and Technology. GALEX is operated for NASA by the California Institute of Technology under NASA contract NAS5-98034.

This paper makes use of MeerKAT observations. The MeerKAT telescope is operated by the South African Radio Astronomy Observatory, which is a facility of the National Research Foundation, an agency of the Department of Science and Innovation.

\end{acknowledgements}

\section*{Data availability}
The HCN and CO data products used in this paper are publicly available via \url{https://www.iram.fr/ILPA/LP015/} (EMPIRE),
\url{https://www.canfar.net/storage/list/phangs/RELEASES/ALMOND/} (ALMOND), and
\url{https://www.canfar.net/storage/list/phangs/RELEASES/PHANGS-ALMA/} (PHANGS--ALMA).
The HCN literature compilation, data products and tables presented in this work are publicly available via \url{https://www.canfar.net/storage/list/phangs/RELEASES/Neumann_etal_2024b/}.
The Python scripts used to create the data products, figures and tables are available via \url{https://github.com/lukas-neumann-astro/publications/tree/main/Neumann_etal_2024b/}.


\bibliographystyle{aa}
\bibliography{references.bib}

\begin{appendix}

\section{EMPIRE versus ALMOND}
\label{sec:app:empire_vs_almond}

\begin{figure}
\centering
\includegraphics[width=\columnwidth]{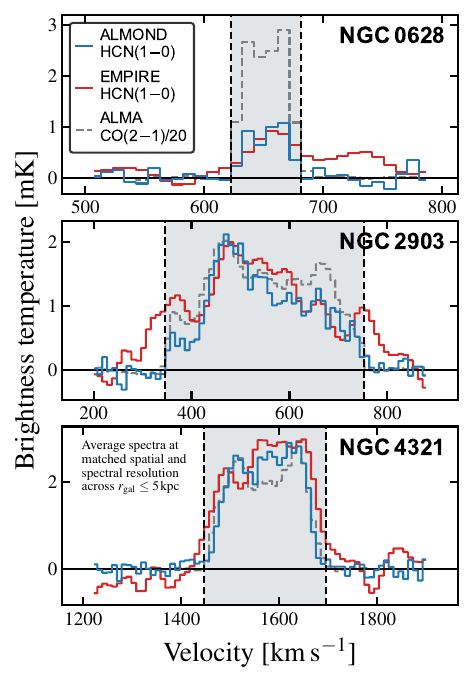}
\caption{EMPIRE versus ALMOND: \hcnone average spectra.
The blue and red lines show average HCN brightness temperatures within $\rgal\leq\SI{5}{\kilo\parsec}$ obtained from spatially and spectrally matched ALMOND and EMPIRE observations, respectively, across the three galaxies NGC\,628, NGC\,2903, NGC\,4321 from top to bottom.
The grey dashed line shows (homogenised) \cotwo intensities from PHANGS--ALMA \citep{Leroy2021b}, scaled down by a factor of 20.
The grey-shaded area indicates the velocity-integration window constructed using the highly significant \cotwo data.
The resulting integrated intensities are quoted in the text.
}
\label{fig:empire_vs_almond_spectra}
\end{figure}

\begin{figure}
\centering
\includegraphics[width=\columnwidth]{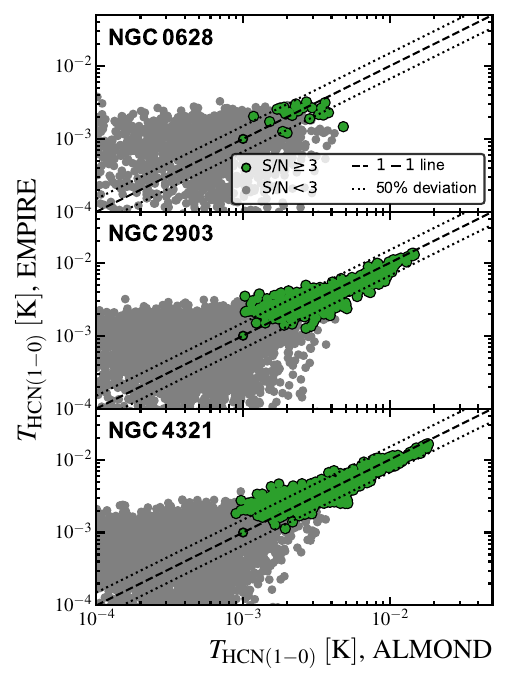}
\caption{EMPIRE versus ALMOND: \hcnone brightness temperature.
Green data points present data, where EMPIRE and ALMOND both yield a 3-sigma detection.
Grey data shows low-significant data points.
The dashed line marks the 1-to-1 relation, where the dotted lines indicate a $\pm\SI{50}{\percent}$ deviation.
}
\label{fig:empire_vs_almond_brightness_temp}
\end{figure}

\begin{figure}
\centering
\includegraphics[width=\columnwidth]{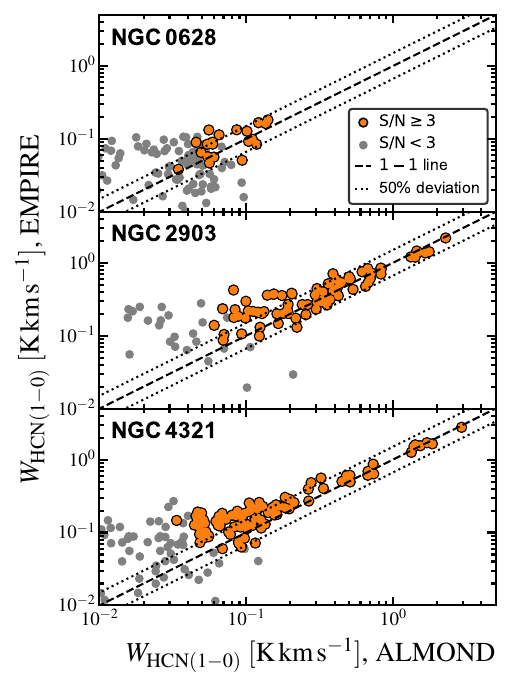}
\caption{EMPIRE versus ALMOND: \hcnone integrated intensity.
Similar to Fig.~\ref{fig:empire_vs_almond_brightness_temp}, but showing the integrated intensities (moment-0) computed across a CO-inferred velocity integration window.
Orange and grey points denote data above and below 3-sigma, respectively.
}
\label{fig:empire_vs_almond_intensity}
\end{figure}

There are three galaxies (i.e. NGC\,628, NGC\,2903, NGC\,4321) that have been mapped in dense gas tracers (e.g. \hcnone) by both surveys, EMPIRE, using the IRAM 30\,m, and ALMOND, using the ACA at similar spectral (a few \SI{}{\km\per\second}) and angular resolution (a few tenths of arcseconds) and sensitivity (a few mK).
In Figs. \ref{fig:empire_vs_almond_spectra} to \ref{fig:empire_vs_almond_intensity} we compare 
the \hcnone data from both surveys.
We homogenise the two datasets by convolving to the best common spectral (i.e. \SI{10}{\km\per\s}) and spatial (i.e. \ang{;;33}) resolution and re-project to the same half-beam size hexagonal pixel grid.

Figure~\ref{fig:empire_vs_almond_spectra} shows average \hcnone spectra computed across all sightlines within \SI{5}{\kilo\parsec} from the galactic centre.
We additional overlay \cotwo average spectra obtained from PHANGS--ALMA \citep{Leroy2021b}, to indicate molecular line emission from a highly significant tracer.
This line has been used to infer the velocity-integration window from which we compute \hcnone integrated intensities of $41.6 \pm \SI{7.3}{\kkms}$, $39.5 \pm \SI{10.7}{\kkms}$ in NGC\,628, $427.8 \pm \SI{47.5}{\kkms}$, $495.7 \pm \SI{143.5}{\kkms}$ in NGC\,2903, and $475.0 \pm \SI{34.6}{\kkms}$, $570.2 \pm \SI{84.8}{\kkms}$ in NGC\,4321 from ALMOND and EMPIRE, respectively.
The average spectra show similar shape and amplitude, demonstrating little to no bias between ALMOND and EMPIRE observations.
The integrated line intensities yield consistent values within their uncertainties.
The largest deviations are observed at large velocity offsets from the galaxies' systemic velocities, potentially linked to poor baseline subtraction.

Figures~\ref{fig:empire_vs_almond_brightness_temp} and \ref{fig:empire_vs_almond_intensity} present a voxel-by-voxel, or pixel-by-pixel comparison between the ALMOND and EMPIRE \hcnone brightness temperatures (ppv cube) and integrated intensities (moment-0 map).
We find that brightness temperatures and integrated intensities agree well between ALMOND and EMPIRE in all galaxies (deviations $\leq\SI{50}{\percent}$ across most detected sightlines) and show little bias ($\leq\SI{10}{\percent}$ on average across all data).
At lower integrated intensities ($\lesssim\SI{e-1}{\kkms}$), EMPIRE yields moderately larger values than ALMOND, which could indicate differences in the calibration and data reduction pipelines.

The comparison between ALMOND and EMPIRE demonstrated that both datasets yield consistent \hcnone intensities and subsequent data products.
In this work, we employ the ALMOND data for the three galaxies NGC\,628, NGC\,2903, NGC\,4321, due to the slightly better angular resolution and sensitivity of the ALMOND survey.

\begin{figure*}
\centering
\includegraphics[width=0.9\textwidth]{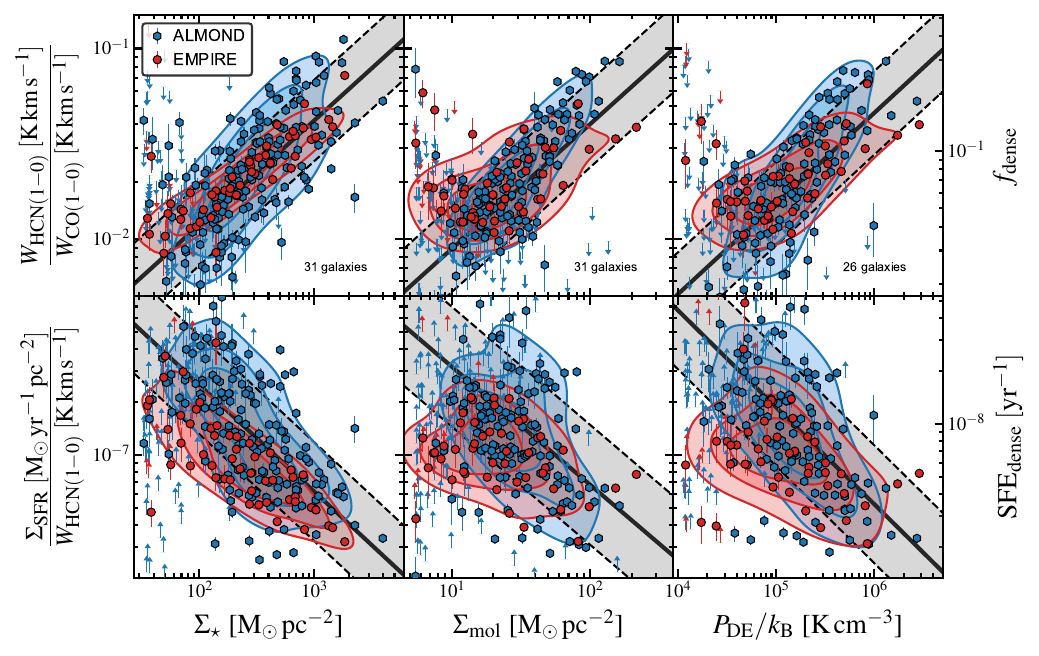}
\caption{Similar to Fig.~\ref{fig:hcn_scaling_relations}, but separately showing kiloparsec-scale, stacked measurements from ALMOND (blue hexagons) and EMPIRE (red circles).
The best-fit line (solid black line) and the corresponding 1-sigma scatter (grey-shaded area) are computed from the combined data using \texttt{LinMix}.
}
\label{fig:hcn_scaling_relations_almond_vs_empire}
\end{figure*}

Figure~\ref{fig:hcn_scaling_relations_almond_vs_empire} shows the HCN/CO and SFR/HCN versus \sigstar, \sigmol, and \pde scaling relations across ALMOND (blue hexagons) and EMPIRE (red circles) at kiloparsec resolution.
Our best-fit relations are similar, though slightly steeper, to those reported by \citet{Jimenez-Donaire2019} but are now measured for a larger and more diverse sample of galaxies. The steeper slopes have two reasons: a) the ALMOND sample shows steeper trends, and (b) the inclusion of non-detections into the fitting routines yields $\sim\SI{10}{\percent}$ steeper slopes. We observe a larger scatter across the full sample of 31 galaxies compared to the nine EMPIRE galaxies alone, suggesting that the more diverse sample captures a wider range of conditions not captured by the simple scaling relations.

\section{Dense gas literature}
\label{sec:app:literature}

In Fig.~\ref{fig:gao_solomon_relation} we present a literature compilation of HCN surveys from local parsec scale over resolved, kiloparsec scale, to unresolved, entire galaxy observations.
The cloud- and clump-scale measurements are taken from observations within the Milky Way \citep{Wu2010, Lada2012, Evans2014, Stephens2016}, the CMZ \citep{Jones2012, Barnes2017}, and the Local Group, namely the Large and Small Magellanic Clouds (LMC, SMC) \citep{Chin1997, Chin1998}, M31 \citep{Brouillet2005}, M33 \citep{Buchbender2013}, and low-metallicity local group galaxies \citep{Braine2017}.
Resolved galaxy observations, typically from nearby galaxies at $100\,$pc to $2\,$kiloparsec scales, include M82 \citep{Kepley2014}, M51 \citep{Usero2015, Chen2017, Querejeta2019, Stuber2023}, NGC\,4038/39 \citep{Bigiel2015}, NGC\,3351, NGC\,3627, NGC\,4254, NGC\,4321, NGC\,5194 \citep{Gallagher2018a}, NGC\,3627 \citep{Beslic2021}, NGC\,1068 \citep{Sanchez-Garcia2022}, NGC\,6946 \citep{Eibensteiner2022}, NGC\,4321 \citep{Neumann2024}, NGC\,253 \citep{Beslic2024}, and the two larger-sample surveys EMPIRE \citep[nine galaxies; ][]{Jimenez-Donaire2019} and ALMOND \citep[25 galaxies; ][]{Neumann2023a}.
Integrated-galaxy data cover Luminous Infrared Galaxies (LIRGs), Ultra-Luminous Infrared Galaxies (ULIRG), and AGN galaxies \citep{Krips2008, Gracia-Carpio2008, Juneau2009, Garcia-Burillo2012, Privon2015}, early-type galaxies \citep{Crocker2012}, and high-redshift galaxies \citep{Gao2007, Rybak2022}.

\section{Conversion factors}
\label{sec:app:conversion_factors}

For EMPIRE, we use the \coone maps obtained as part of the survey. 
For ALMOND, we use PHANGS--ALMA \cotwo maps, which we convert to an equivalent \coone intensity before applying $\alphaco$.
To do this, we estimate a line ratio, $R_{21}$, based on the local SFR surface density (\sigsfr) following \citet{denBrok2021}, \citet{Leroy2022}, and \citet{Schinnerer2024}:
\begin{align}
    R_{21} = \frac{\cotwo}{\coone} = 0.65 \,  \left(\frac{\sigsfr}{\SI{1.8e-2}{\msun\per\year\per\square\kilo\parsec}}\right)^{0.125}  \;,
\end{align}
with minimum $R_{21}$ of 0.35 and maximum 1.0.
Then we scale the \cotwo intensity by $R_{21}^{-1}$ to present our results in terms of \coone intensity.

To compute \sigmol and \pde, we adopt the variable \alphaco prescription from \citet[][their Table~1]{Schinnerer2024}, which accounts for variations with metallicity ($Z$; \SI{}{\zsun} is the solar metallicity) and stellar mass surface density (\sigstar):
\begin{align}
    \alphaco^{\rm var} = \alphaco^{\rm fix} \, \left(\dfrac{Z}{\SI{}{\zsun}}\right)^{-1.5} \, \left(\dfrac{\mathrm{max}(\sigstar,\,\SI{100}{\msun\per\square\parsec})}{\SI{100}{\msun\per\square\parsec}}\right)^{-0.25} \;.
\end{align}
Stellar mass maps are inferred from \textit{Spitzer} \SI{3.6}{\micron} observations as explained in Sect.~\ref{sec:data}.
Metallicities are estimated based on simple scaling relations, following \citet{Sun2020a}.
These use a global mass-metallicity relation \citep{Sanchez2019} and employ a radial metallicity relation with a fixed gradient of \SI{-0.1}{\dex} normalised by the effective radius of each galaxy \citep{Sanchez2014}.

\section{Spectral stacking of HCN and CO}
\label{sec:app:stacking}

\begin{figure*}
\centering
\includegraphics[width=0.9\textwidth]{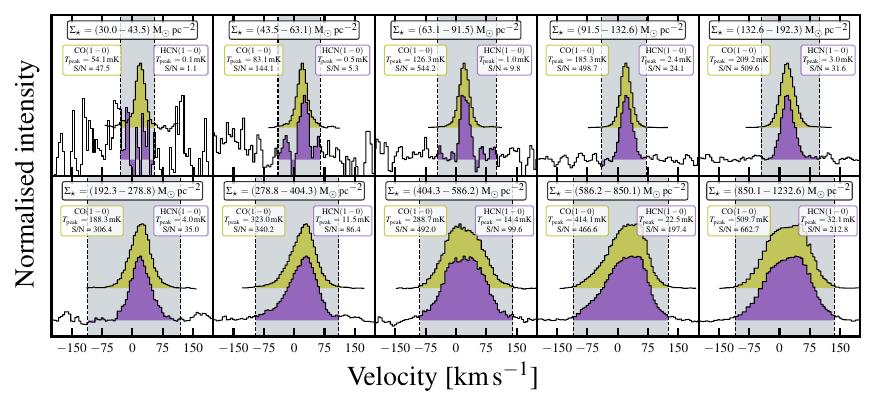}
\caption{Spectral stacks of CO (olive) and HCN (purple) across NGC\,4321 in logarithmically-spaced bins of \sigstar. The grey-shaded area indicates the velocity-integration window applied to compute the integrated intensities of the stacked spectra. The labelled boxes show the peak intensity and \snr of the integrated intensities of CO and HCN, respectively, for each stacked spectrum.}
\label{fig:ngc4321_stacks}
\end{figure*}

We compute spectral stacks of \hcnone and \coone line emission in bins of stellar mass surface density, \sigstar, molecular gas surface density, \sigmol, and dynamical equilibrium pressure, \pde, across the merged sample of 31 galaxies studied in this work.
For the ALMOND sample, the \cotwo intensities from PHANGS--ALMA are first converted into \coone intensities using the line ratio calibration from Sect.~\ref{sec:app:conversion_factors}. 
We note, however, that this has no effect on the stacking procedure. 
We stack in logarithmic bins for each galaxy individually, selecting ten bins from a fixed minimum ($\sigstar=\SI{3e1}{\msun\per\square\parsec}$, $\sigmol=\SI{5e0}{\msun\per\square\parsec}$, $\pde=\SI{1e4}{\kB\kelvin\per\cubic\cm}$) to the maximum value of each galaxy.
For the centres versus disc HCN scaling relations (Fig.~\ref{fig:hcn_scaling_relations}), we exclude the centres from the stacking and stacks across the remaining sightlines adopting nine bins.
Figure~\ref{fig:ngc4321_stacks} shows exemplary spectral stacks of \hcnone and \coone as a function of \sigstar across the galaxy NGC\,4321.

\section{Additional tables}
\label{sec:app:fig_tab}

Table~\ref{tab:sample} presents the combined galaxy sample composed of 31 galaxies from the EMPIRE and ALMOND surveys, along with their coordinates and global properties.

\begin{table*}
\begin{center}
\caption{Galaxy sample (EMPIRE + ALMOND).}
\label{tab:sample}
\resizebox{\textwidth}{!}{
\begin{tabular}{cccccccccccccccc}
    \hline\hline
    \multirow{2}{*}{Galaxy} & R.A. & Dec. & $d$ & $i$ & $\log_{10}\,M_\star$ & $\log_{10}\,$SFR & $\log_{10}\,(\mathrm{SFR}/M_\star)$ & \multirow{2}{*}{Bar} & \multirow{2}{*}{AGN} & \multirow{2}{*}{SFR tracer} & \multirow{2}{*}{\hone survey} & \multirow{2}{*}{CO survey} & \multirow{2}{*}{HCN survey} & \multicolumn{2}{c}{HCN resolution} \\ 
    & (J2000) & (J2000) & ($\SI{}{\mega\parsec}$) & ($\SI{}{\degree}$) & ($\SI{}{\msun}$) & ($\SI{}{\msun\per\year}$) & ($\SI{}{\per\year}$) & & & & & & & ($\SI{}{\arcsecond}$) & (kpc) \\
    (1) & (2) & (3) & (4) & (5) & (6) & (7) & (8) & (9) & (10) & (11) & (12) & (13) & (14) & (15) & (16) \\
    \hline
NGC\,0628 & \ra{01;36;41.7252} & \ang{+15;47;01.1148} & $9.84$ & $8.90$ & $10.34$ & $0.24$ & $-10.10$ & \xmark & \xmark & W4+FUV & THINGS & PHANGS-ALMA & ALMOND & $18.60$ & $0.89$ \\
NGC\,1097 & \ra{02;46;18.94968} & \ang{-30;16;28.83} & $13.58$ & $48.60$ & $10.76$ & $0.68$ & $-10.08$ & \cmark & \cmark & W4+FUV & PHANGS-MeerKAT & PHANGS-ALMA & ALMOND & $19.40$ & $1.28$ \\
NGC\,1365 & \ra{03;33;36.3648} & \ang{-36;08;25.4544} & $19.57$ & $55.40$ & $10.99$ & $1.23$ & $-9.76$ & \cmark & \cmark & W4+FUV & \xmark & PHANGS-ALMA & ALMOND & $20.60$ & $1.95$ \\
NGC\,1385 & \ra{03;37;28.5636} & \ang{-24;30;04.1832} & $17.22$ & $44.00$ & $9.98$ & $0.32$ & $-9.66$ & \xmark & \xmark & W4+FUV & PHANGS-VLA & PHANGS-ALMA & ALMOND & $19.90$ & $1.66$ \\
NGC\,1511 & \ra{03;59;36.5904} & \ang{-67;38;02.148} & $15.28$ & $72.70$ & $9.91$ & $0.36$ & $-9.55$ & \xmark & \xmark & W4+FUV & PHANGS-MeerKAT & PHANGS-ALMA & ALMOND & $17.60$ & $1.30$ \\
NGC\,1546 & \ra{04;14;36.2928} & \ang{-56;03;39.2328} & $17.69$ & $70.30$ & $10.35$ & $-0.08$ & $-10.43$ & \xmark & \xmark & W4+FUV & \xmark & PHANGS-ALMA & ALMOND & $18.90$ & $1.62$ \\
NGC\,1566 & \ra{04;20;00.3816} & \ang{-54;56;16.836} & $17.69$ & $29.50$ & $10.78$ & $0.66$ & $-10.13$ & \cmark & \cmark & W4+FUV & MHONGOOSE & PHANGS-ALMA & ALMOND & $19.70$ & $1.69$ \\
NGC\,1672 & \ra{04;45;42.4896} & \ang{-59;14;50.1252} & $19.40$ & $42.60$ & $10.73$ & $0.88$ & $-9.85$ & \cmark & \cmark & W4+FUV & MHONGOOSE & PHANGS-ALMA & ALMOND & $18.20$ & $1.71$ \\
NGC\,1792 & \ra{05;05;14.3256} & \ang{-37;58;50.016} & $16.20$ & $65.10$ & $10.61$ & $0.57$ & $-10.04$ & \xmark & \xmark & W4+FUV & \xmark & PHANGS-ALMA & ALMOND & $18.70$ & $1.47$ \\
NGC\,2566 & \ra{08;18;45.6072} & \ang{-25;29;58.272} & $23.44$ & $48.50$ & $10.71$ & $0.94$ & $-9.77$ & \cmark & \xmark & W4 & PHANGS-VLA & PHANGS-ALMA & ALMOND & $18.50$ & $2.10$ \\
NGC\,2903 & \ra{09;32;10.1064} & \ang{+21;30;03.0276} & $10.00$ & $66.80$ & $10.63$ & $0.49$ & $-10.15$ & \cmark & \xmark & W4+FUV & THINGS & PHANGS-ALMA & ALMOND & $18.30$ & $0.89$ \\
NGC\,2997 & \ra{09;45;38.7936} & \ang{-31;11;27.924} & $14.06$ & $33.00$ & $10.73$ & $0.64$ & $-10.09$ & \xmark & \xmark & W4+FUV & PHANGS-VLA & PHANGS-ALMA & ALMOND & $20.40$ & $1.39$ \\
NGC\,3059 & \ra{09;50;08.16} & \ang{-73;55;19.902} & $20.23$ & $29.40$ & $10.38$ & $0.38$ & $-10.00$ & \cmark & \xmark & W4 & PHANGS-MeerKAT & PHANGS-ALMA & ALMOND & $16.70$ & $1.64$ \\
NGC\,3184 & \ra{10;18;16.9416} & \ang{+41;25;27.6348} & $12.58$ & $16.00$ & $10.36$ & $0.11$ & $-10.24$ & \cmark & \cmark & W4+FUV & THINGS & EMPIRE & EMPIRE & $33.30$ & $2.03$ \\
NGC\,3521 & \ra{11;05;48.576} & \ang{-00;02;09.4164} & $13.24$ & $68.80$ & $11.02$ & $0.57$ & $-10.45$ & \xmark & \xmark & W4 & THINGS & PHANGS-ALMA & ALMOND & $21.10$ & $1.35$ \\
NGC\,3621 & \ra{11;18;16.3008} & \ang{-32;48;45.36} & $7.06$ & $65.80$ & $10.06$ & $-0.00$ & $-10.06$ & \xmark & \cmark & W4+FUV & THINGS & PHANGS-ALMA & ALMOND & $18.90$ & $0.65$ \\
NGC\,3627 & \ra{11;20;15.0048} & \ang{+12;59;29.4} & $11.32$ & $57.30$ & $10.83$ & $0.58$ & $-10.25$ & \cmark & \cmark & W4+FUV & THINGS & EMPIRE & EMPIRE & $33.30$ & $1.83$ \\
NGC\,4254 & \ra{12;18;49.632} & \ang{+14;24;59.0832} & $13.10$ & $34.40$ & $10.42$ & $0.49$ & $-9.94$ & \xmark & \xmark & W4+FUV & VLA-HERACLES & EMPIRE & EMPIRE & $33.30$ & $2.11$ \\
NGC\,4303 & \ra{12;21;54.9312} & \ang{+04;28;25.4784} & $16.99$ & $23.50$ & $10.52$ & $0.73$ & $-9.80$ & \cmark & \cmark & W4+FUV & PHANGS-VLA & PHANGS-ALMA & ALMOND & $20.20$ & $1.66$ \\
NGC\,4321 & \ra{12;22;54.9288} & \ang{+15;49;20.2944} & $15.21$ & $38.50$ & $10.75$ & $0.55$ & $-10.19$ & \cmark & \xmark & W4+FUV & VLA-HERACLES & PHANGS-ALMA & ALMOND & $19.60$ & $1.45$ \\
NGC\,4535 & \ra{12;34;20.304} & \ang{+08;11;52.7028} & $15.77$ & $44.70$ & $10.53$ & $0.33$ & $-10.20$ & \cmark & \xmark & W4+FUV & PHANGS-MeerKAT & PHANGS-ALMA & ALMOND & $22.80$ & $1.74$ \\
NGC\,4536 & \ra{12;34;27.0672} & \ang{+02;11;17.6748} & $16.25$ & $66.00$ & $10.40$ & $0.54$ & $-9.86$ & \cmark & \xmark & W4+FUV & VLA-HERACLES & PHANGS-ALMA & ALMOND & $21.50$ & $1.69$ \\
NGC\,4569 & \ra{12;36;49.824} & \ang{+13;09;46.35} & $15.76$ & $70.00$ & $10.81$ & $0.12$ & $-10.68$ & \cmark & \cmark & W4+FUV & VIVA & PHANGS-ALMA & ALMOND & $19.20$ & $1.47$ \\
NGC\,4826 & \ra{12;56;43.6416} & \ang{+21;40;59.0988} & $4.41$ & $59.10$ & $10.24$ & $-0.69$ & $-10.93$ & \xmark & \cmark & W4+FUV & THINGS & PHANGS-ALMA & ALMOND & $18.70$ & $0.40$ \\
NGC\,5055 & \ra{13;15;49.296} & \ang{+42;01;45.4008} & $9.02$ & $59.00$ & $10.79$ & $0.31$ & $-10.48$ & \xmark & \xmark & W4+FUV & THINGS & EMPIRE & EMPIRE & $33.30$ & $1.46$ \\
NGC\,5194 & \ra{13;29;52.6896} & \ang{+47;11;42.5472} & $8.56$ & $21.00$ & $10.65$ & $0.64$ & $-10.01$ & \xmark & \cmark & W4+FUV & THINGS & PAWS & EMPIRE & $33.30$ & $1.38$ \\
NGC\,5248 & \ra{13;37;32.0064} & \ang{+08;53;06.702} & $14.87$ & $47.40$ & $10.41$ & $0.36$ & $-10.05$ & \cmark & \xmark & W4+FUV & PHANGS-VLA & PHANGS-ALMA & ALMOND & $19.90$ & $1.43$ \\
NGC\,5643 & \ra{14;32;40.776} & \ang{-44;10;28.596} & $12.68$ & $29.90$ & $10.34$ & $0.41$ & $-9.92$ & \cmark & \cmark & W4 & \xmark & PHANGS-ALMA & ALMOND & $18.00$ & $1.11$ \\
NGC\,6300 & \ra{17;16;59.472} & \ang{-62;49;13.98} & $11.58$ & $49.60$ & $10.47$ & $0.28$ & $-10.19$ & \cmark & \cmark & W4 & \xmark & PHANGS-ALMA & ALMOND & $17.70$ & $0.99$ \\
NGC\,6946 & \ra{20;34;52.6032} & \ang{+60;09;12.654} & $7.34$ & $33.00$ & $10.47$ & $0.77$ & $-9.70$ & \cmark & \xmark & W4+FUV & THINGS & EMPIRE & EMPIRE & $33.30$ & $1.18$ \\
NGC\,7496 & \ra{23;09;47.2848} & \ang{-43;25;40.26} & $18.72$ & $35.90$ & $10.00$ & $0.35$ & $-9.64$ & \cmark & \cmark & W4+FUV & PHANGS-MeerKAT & PHANGS-ALMA & ALMOND & $17.90$ & $1.62$ \\
    \hline\hline
\end{tabular}
}
\end{center}
\footnotesize{
    \textbf{Notes} -- (2) Right ascension, (3) declination, (4) distance \citep{Anand2021}, and (5) inclination angle \citep{Lang2020}.
    Integrated galaxy properties, (6) global stellar mass and (7) global SFR, taken from \cite{Leroy2019}.
    (9) Presence of a galactic bar \citep{Herrera-Endoqui2015, Querejeta2021b}, and/or (10) AGN \citep{Veron2010}.
    (11) Employed SFR tracers using WISE \SI{22}{\micron} \citep{Wright2010} or a combination of WISE and GALEX-FUV \citep{Martin2005}, aopted from \citep{Leroy2019}.
    (12) Archival \hone 21-cm line emission data, taken from THINGS \citep{Walter2008}, VIVA \citep{Chung2009}, VLA--HERACLES, PHANGS--VLA (Utomo et al. in prep.), PHANGS--MeerKAT \citep[][; Pisano et al. in prep.]{Eibensteiner2024}, and MHONGOOSE \citep{deBlok2024}.
    (13) Archival CO observations from PHANGS--ALMA \citep[\cotwo;][]{Leroy2021b} for ALMOND galaxies, and EMPIRE \citep[\coone;][]{Jimenez-Donaire2019}, PAWS \citep[\coone;][]{Schinnerer2013} for EMPIRE galaxies.
    (14) Adopted HCN data from ALMOND \citep{Neumann2023a} and EMPIRE \citep{Jimenez-Donaire2019}. 
    (15) HCN native angular resolution and (16) corresponding linear resolution, given the distance $d$.
}
\end{table*}

Table~\ref{tab:centres} lists percentile and median HCN/CO and SFR/HCN values for centre and discs environments discussed in Sect.~\ref{sec:centres}.

\begin{table*}
\begin{center}
\caption{Galaxy centres vs discs.}
\label{tab:centres}
\begin{tabular}{cccc}
\hline \hline
$\log_{10}(Y)$ &  environment & (16$^\mathrm{th}$, 50$^\mathrm{th}$, 84$^\mathrm{th}$) perc. ($\snr\geq 3$) & median (all \snr) \\
\hline
\multirow{4}{*}{HCN/CO}  &         disc     & $(-1.96,-1.72,-1.51)$ & $-1.87$ \\
                         &       centre     & $(-1.60,-1.35,-1.04)$ & $-1.35$ \\
                         & centre (AGN)     & $(-1.43,-1.30,-1.03)$ & $-1.28$ \\
                         & centre (non-AGN) & $(-1.62,-1.43,-1.22)$ & $-1.43$ \\
\hline
\multirow{4}{*}{SFR/HCN} &         disc     & $(-7.21,-6.93,-6.56)$ & $-6.85$ \\
                         &       centre     & $(-7.39,-7.13,-6.57)$ & $-7.11$ \\
                         & centre (AGN)     & $(-7.44,-7.19,-6.70)$ & $-7.23$ \\
                         & centre (non-AGN) & $(-7.31,-7.11,-6.57)$ & $-7.11$ \\
                         
\hline\hline
\end{tabular}
\end{center}
{\bf Notes} --
16$^\mathrm{th}$ percentiles, medians, and 84$^\mathrm{th}$ percentiles of HCN/CO and SFR/HCN across significant data ($\snr\geq 3$) in disc and centre environments.
For the centres, we also present percentile values among centres with and without AGNs.
The last column additionally lists the median values across all \snr data, this means including non-detections, which yields \SI{0.15}{\dex} lower HCN/CO and \SI{0.08}{\dex} higher SFR/HCN across galaxy discs, which are affected by \snr-clipping (and similar values in centres, which have $\snr\geq 3$ for 29 out of 31 galaxies).
\end{table*}

\end{appendix}

\end{document}